\documentclass{article}
\usepackage{arxiv}
\usepackage[utf8]{inputenc} 
\usepackage[T1]{fontenc}    
\usepackage{hyperref}       
\usepackage{url}            
\usepackage{booktabs}       
\usepackage{amsfonts}       
\usepackage{nicefrac}       
\usepackage{microtype}      
\usepackage{multicol}
\usepackage{graphicx}
\graphicspath{ {./images/} }
\usepackage{listings, xcolor}

\definecolor{verylightgray}{rgb}{.97,.97,.97}

\lstdefinelanguage{Solidity}{
	keywords=[1]{anonymous, assembly, assert, balance, break, call, callcode, case, catch, class, constant, continue, constructor, contract, debugger, default, delegatecall, delete, do, else, emit, event, experimental, export, external, false, finally, for, function, gas, if, implements, import, in, indexed, instanceof, interface, internal, is, length, library, log0, log1, log2, log3, log4, memory, modifier, new, payable, pragma, private, protected, public, pure, push, require, return, returns, revert, selfdestruct, send, solidity, storage, struct, suicide, super, switch, then, this, throw, transfer, true, try, typeof, using, value, view, while, with, addmod, ecrecover, keccak256, mulmod, ripemd160, sha256, sha3}, 
	keywordstyle=[1]\color{blue}\bfseries,
	keywords=[2]{address, bool, byte, bytes, bytes1, bytes2, bytes3, bytes4, bytes5, bytes6, bytes7, bytes8, bytes9, bytes10, bytes11, bytes12, bytes13, bytes14, bytes15, bytes16, bytes17, bytes18, bytes19, bytes20, bytes21, bytes22, bytes23, bytes24, bytes25, bytes26, bytes27, bytes28, bytes29, bytes30, bytes31, bytes32, enum, int, int8, int16, int24, int32, int40, int48, int56, int64, int72, int80, int88, int96, int104, int112, int120, int128, int136, int144, int152, int160, int168, int176, int184, int192, int200, int208, int216, int224, int232, int240, int248, int256, mapping, string, uint, uint8, uint16, uint24, uint32, uint40, uint48, uint56, uint64, uint72, uint80, uint88, uint96, uint104, uint112, uint120, uint128, uint136, uint144, uint152, uint160, uint168, uint176, uint184, uint192, uint200, uint208, uint216, uint224, uint232, uint240, uint248, uint256, var, void, ether, finney, szabo, wei, days, hours, minutes, seconds, weeks, years},	
	keywordstyle=[2]\color{teal}\bfseries,
	keywords=[3]{block, blockhash, coinbase, difficulty, gaslimit, number, timestamp, msg, data, gas, sender, sig, value, now, tx, gasprice, origin},	
	keywordstyle=[3]\color{violet}\bfseries,
	identifierstyle=\color{black},
	sensitive=false,
	comment=[l]{//},
	morecomment=[s]{/*}{*/},
	commentstyle=\color{gray}\ttfamily,
	stringstyle=\color{red}\ttfamily,
	morestring=[b]',
	morestring=[b]"
}

\title{Enhancing Healthcare System Using Blockchain Smart Contracts}

\author{
 Shashank Joshi \\
 Department of Computer Science and Engineering\\
 SRM Institute Of Science And Technology\\
 Kattankulathur, Tamil Nadu – 603203, \\
 \texttt{sj8559@srmist.edu.in} \\
   \And
Arhan Choudhury \\
 Department of Computer Science and Engineering\\
 SRM Institute Of Science And Technology\\
 Kattankulathur, Tamil Nadu – 603203, \\
 \texttt{ac8365@srmist.edu.in} \\
  \And
 Ojas Saraswat \\
 Department of Computer Science and Engineering\\
 SRM Institute Of Science And Technology\\
 Kattankulathur, Tamil Nadu – 603203, \\
 \texttt{os5373@srmist.edu.in} \\
}

\begin{document}
\maketitle

\begin{abstract}
\textbf{The concept of blockchain has emerged as an effective solution for data-sensitive domains, such as healthcare, financial services, etc., due to its various attributes like immutability, non-repudiation, and availability. Thus, implementation of this technology in various domains rose exponentially; one of such fields is the healthcare supply chain. Managing healthcare supply chain processes effectively is very crucial for the healthcare system. Despite various innovations in the method of treatment methodologies, the healthcare supply chain management system is not up to the mark and lacks efficiency. The traditional healthcare supply chain system is time-consuming and lacks the work synergy among the various stakeholders of the supply chain. Thus, In this paper, we propose a framework based on blockchain smart contracts and decentralized storage to connect all the supply chain stakeholders. Smart contracts in the framework enforce and depict various interactions and transactions among the stakeholders, thus helping to automate these processes, promote transparency, improve efficiency, and minimize service time. The preliminary results show that the proposed framework is more efficient, secure, and economically feasible.
}
\end{abstract}


\section{Introduction}

With the development of cryptocurrencies like bitcoin \cite{1}, blockchain emerged as a propitious technology that can be applied to solve problems of several data-sensitive domains. A blockchain is a decentralized and immutable growing list of records integrated with cryptographic protocols \cite{2} and has salient properties such as resilience, non-repudiation, and tamper resistance. In the context of the healthcare system, the development in healthcare practices have improved medical quality and efficiency \cite{3} but still there are some problems regarding the accessibility of healthcare services and proper handling, data sharing, and storage of medical records and the integration of blockchain can provide a significant improvement to enhance the patient service quality and safety \cite{4}, it can also streamline all the healthcare processes thus helps to create Patient-centric systems which require trust, transparency and effective information sharing among various stakeholders of the system \cite{5} and foster an era of growth and innovation.\\
The objective of this paper is to present and evaluate the use of blockchain technology as a service to enhance the healthcare system and streamline the healthcare services and supply chain. The core of this paper is the use of a blockchain-based smart contract, a computer program, or transaction protocol to digitally facilitate, verify, or enforce the transaction (or interaction) and imitate a real-world contract. Smart contracts are instrumental in reducing arbitration, enforcement costs, fraud losses, and cognitive overload on healthcare services \cite{6}. Understanding the current state of healthcare problems, reducing the costs of healthcare services, increasing resource utilization, and evaluating the practical feasibility of smart-contract is the focal objective of this paper.\\  
The remainder of this paper is organized as follows: Section 2  presents some background. Related work is presented in Section 3. Section 4 discusses design considerations for healthcare systems. Section 5 presents the proposed blockchain-based healthcare system. Finally, Section 6 concludes this paper and indicates some future work.
\section{Background}
Blockchain is a decentralized, immutable ledger on a peer to peer network, and its decentralized nature triggers complexity to validate and verify the transactions \cite{2}, as there is no central administration or centralized database \cite{7}, to overcome this complexity a consensus algorithm \cite{8} is introduced which is an instrument using which nodes in the blockchain can come to single truth state without a central authority. Thus the consensus algorithm ensures the reliability and integrity of data in this untrusted environment. The blockchain network only allows to append the blocks, and the hash of the previous block is stored in the successor blocks of the chain, it guarantees immutability, as the data of pre-existing blocks cannot be altered or deleted.\\
In general, blockchain networks can be divided into private, public, and consortium-based. A private blockchain is a type of blockchain network which can limit the read and write access, and can specify the node that can validate and verify the transactions. Thus, the transactions on private blockchain networks are cheaper and have shorter block times \cite{9}. Public blockchains allow any user or node on that network to read and create a transaction. These blockchain networks are permissionless in nature, and anyone can participate in the network. A consortium blockchain is a partially decentralized blockchain network where a set of nodes are authorized and responsible for the consensus in the blockchain network.\\ 
Depending upon the application, blockchain has different characteristics; thus, to integrate the business logic with the blockchain network, smart contracts are used. Smart contracts are trackable and irreversible scripts representing a real-world contract that can be enforced automatically, reducing the need for intermediaries \cite{10} in a decentralized environment. Smart Contract provides flexibility to process any application, perform the required business logic or operation, provide immutability of the generated data, transparency, and auditability on performed processes or transactions.\\
Each Ethereum node has a virtual machine called Ethereum Virtual Machine, which acts like a decentralized computer and is responsible for processing bytecodes representing smart contracts\cite{11}. It is the runtime environment for the smart contracts associated with the blockchain network. Peers in the network make requests by calling smart contracts, enabling them to change their state and return information regarding the current state. Blockchain nodes process these requests on the smart contract bytecode in its EVM and store the result in the blockchain \cite{12}.
\section{Related Work}
Different research proposed methods for solving the healthcare services issues and the adoption of blockchain-based smart contracts. For example, Daisuke et al. \cite{13} examine the storage of medical records to the hyper ledger blockchain network using the mobile device. In their work, they tried to make the data stored on the network tamperproof.\\
Shen et al. \cite{14} propose a blockchain-based framework known as MedChain for the effective sharing of medical data generated by the medical examination and the patient data collected from various IoT devices.\\
Jamil et al. \cite{15} discussed the issues regarding the integrity in the drug supply chain and proposed the use of blockchain for the standardization of the drugs and to detect counterfeits.\\
Anuraag et al. \cite{16} came up with a blockchain-based model to manage medical information. Their study reviews all the benefits and drawbacks of blockchain technology for the healthcare sector.\\
Litchfield et al. \cite{17} have highlighted and surveyed security and privacy issues associated with healthcare data and proposed the use of blockchain to overcome these issues.\\  
Rouhani et al. \cite{18} examine the security issues and performance of permissioned and permissionless blockchain networks based on an instance of a patient-controlled healthcare management system.\\
Kumar et al.\cite{19} presented various blockchain applications for the healthcare system and highlighted various limitations in the integration of blockchain technology with the healthcare management system.\\
Vora et al.\cite{20} proposed a blockchain mechanism to handle and store electronic health records efficiently. The core objective of this study was to analyze how the proposed framework maintains privacy and security.
\section{Design And Implementation}
For the development of the proposed solution, a set of technologies are used. The framework implementation of the solution will be a decentralized application with a distributed file system. The solution development can be divided into three levels:
\begin{enumerate}
  \item \textbf{Implementation of backend}\\
Smart contracts are used for the implementation of the backend. Smart contracts comprise events, functions, state variables, and modifiers and are developed in solidity, the main language on Ethereum.
  \item \textbf{Application Programming Interface(API) and technologies used to communicate between various levels}\\
Infura's API provides instant and reliable HTTPS and WebSocket access to the Ethereum and IPFS networks, and Web3 is used for establishing a connection with the blockchain network. 
  \item \textbf{Frontend}\\ 
The React.js libraries are used for the implementation of the web interface along with the Web3.js (based on web3 Ethereum interface) to establish the connection with the blockchain network.
\end{enumerate}
\section{Proposed Solution And Architecture}
In this paper, we consider the existing constraints and problems in the healthcare system and supply chain like data protection, information sharing among various stakeholders, streamlining services, etc. Based on this, a blockchain-based healthcare system solution is proposed for optimizing the identified requirements and considerations. The solution devised is a blockchain-based smart contract for the healthcare system. In the following subsections, we start by explaining and identifying the roles followed by various processes required for implementing healthcare system smart contracts. In the last subsection, we analyze the security of the proposed system.
\subsection{Roles and Stakeholders}
As represented in Figure 1, the proposed blockchain-based healthcare system allows the participation of individuals or institutions in the following roles.
\begin{enumerate}
  \item \textbf{Administrator: }It is responsible for managing and maintaining the healthcare network. This role can be assigned to multiple trusted institutions and service points. The administrator is responsible for registering other stakeholders in the network and also exercises the right to remove a stakeholder from the network in given circumstances and assign permissioned nodes, thus enforcing the prescribed guidelines.
  \item \textbf{Patients: }Patients are the set of users in the network who can be any recipient of healthcare and allied services. Patients can interact with other stakeholders to perform a variety of operations like accessing their medical records, getting prescriptions, raising insurance claims, and ordering the prescribed drugs and medicines.
  \item \textbf{Doctor: }Set of registered medical practitioners associated with a hospital and are capable of examining the patient's medical record, adding prescription to the patient's medical record, and examining and verifying the patient's medical data for insurance companies.
  \item \textbf{Hospital: }The institutes which provide health care services to the patients and are capable of adding medical records to patient record list and book appointments. 
  \item \textbf{Insurance Company: }The companies which provide the insurance-related services and are capable of examining the medical expenses and bills of the registered patient for their medical claim and disperse the funds.
  \item \textbf{Pharmacy: }The entity which provides medications and drugs based on the patient's prescription and is capable of assigning the medical bills in the patient's medical record.  
\end{enumerate}
\begin{figure} 
    \centering
    \begin{center}
    \includegraphics{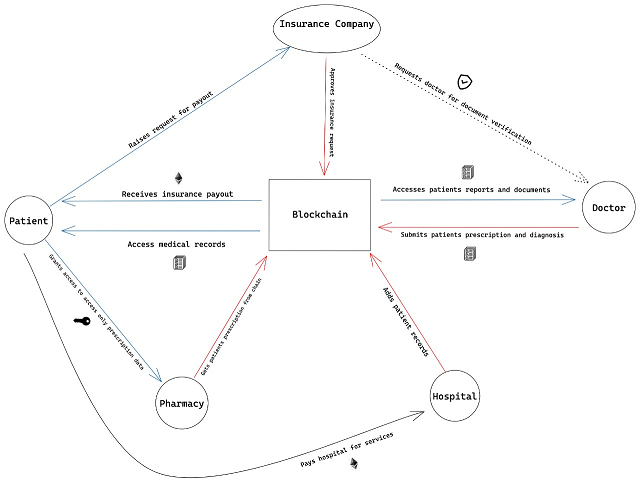}
    \end{center}
    \caption{Roles And Processes}
\end{figure}
\subsection{Healthcare Processes}
In our proposed solution, the healthcare services and processes are represented by the smart contract instantiated on the blockchain by the administrators. The following are the significant activities in the healthcare process: 
\begin{enumerate}
  \item \textbf{Stakeholders Registration: }The phase of registration of various stakeholders is conducted by administrators or permissioned nodes in the network. This requires the verification of individuals and institutions based on the government guidelines, which requires a component for a government verification service. For each successful registration, a corresponding wallet address is generated.
  \item \textbf{Issuing And Filing Medical Record: }To streamline the medical service and eliminate its constraints, proper handling of the data is very necessary. Thus, In our proposed medical records are mapped with the patient's address and the bills and prescriptions are stored in the form of an IPFS hash on the blockchain network. In this phase, the hospital creates a new medical record for the patient with the help of all the relevant information, then during the medical checkup, a doctor writes a prescription for the patient and appends it in the patient's medical record in the form of an IPFS hash value through the smart contract. The smart contract features a smooth data flow between various stakeholders and organizes the patient’s medical history.
  \item \textbf{Accessing Information: }In the proposed model, the medical data can be accessed in different ways at different levels and interaction of stakeholders. In patient-hospital interaction, only the relevant information regarding the patient's current medical appointment is accessed by the hospital. On the other hand, a doctor can refer to the patient's past medical history during the medical checkup. While the pharmacy and insurance companies can only access the data when authorized to do so by the patient.
  \item \textbf{Effective Data Sharing between Patients and Pharmacy: }In this phase, a patient generates a request for a medical service like pharmacy, where the patient authorizes and shares a specific medical prescription with the pharmacy. After accessing the prescription, the pharmacy then issues the required medicines and appends the bills in the patient’s medical record as a hash through a smart contract. After this step, the permissions given by the patient are reset. This mechanism is the same for all other health service providers like laboratories etc.
  \item \textbf{Healthcare reimbursement: }In this phase, the patient with a health insurance policy raises a request for payout against a specific medical record which contains all the bills and prescription; the health insurance company can examine and verify the medical record and all the associated documents and initiate a reimbursement to the patient. After the payment, the request status is reset to its initial state.
  \item \textbf{Transactions: } As the proposed model is based on the blockchain architecture, all the transactions among various stakeholders will be efficient, secure, and transparent.  
\end{enumerate}
\subsection{Index of Functionalities}
In this subsection, we will elaborate the functionality of the healthcare system smart contract.
\subsubsection{Patient Functionalities}
\begin{lstlisting}[language=Solidity]
function getrecord(uint _rid) public view checkpe(msg.sender) recexist(msg.sender, _rid) returns(address _d, uint256 adm, uint256 dis, string memory pres, string memory bill){
        return (
           patient_list[msg.sender].recordlist[_rid].doctor,
           patient_list[msg.sender].recordlist[_rid].admissionDate,
           patient_list[msg.sender].recordlist[_rid].dischargeDate,
           patient_list[msg.sender].recordlist[_rid].pres,
           patient_list[msg.sender].recordlist[_rid].bill
        );
 }
 function getrecordnum () public view checkpe(msg.sender) returns(uint256 _rid){
     return patient_list[msg.sender].recn;
 }
 function triggerpayment(address _b) public payable {
     (bool status,)=_b.call{value:msg.value}("");
      require(status,"Transaction Unsucessful");

 }
function allowph (address _b, uint256 _rid) public recexist(msg.sender, _rid){
    permlist[msg.sender][_b].recid=_rid;
    permlist[msg.sender][_b].all=true;

}
function allowi (address _b, uint256 _rid) public allowins(msg.sender){
    ilist[msg.sender][_b]=_rid;
    insurancec_list[_b].flagraised[msg.sender]=true;
}
function getdoctor(address _d) public view checkde(_d) returns(string memory _name, address _hospital, string memory _spec)
{
        return(
            doctor_list[_d].name,
            doctor_list[_d].hos,
            doctor_list[_d].spec

        );
}   
\end{lstlisting}
\begin{itemize}
\item getrecord: Takes the record number and returns the respective medical record of the patient.
\item getrecordnum: Returns an unsigned integer representing the total number of medical records associated with a patient.
\item triggerpayment: Takes the address of the beneficiary and sends the designated amount of tokens(eth) as a payment from the patient's account.
\item allowph: Takes the address of the pharmacy and the record number and allows the pharmacy to access the prescription associated with that medical record.
\item allowi: Takes the address of the insurance company and the record number and raises the request for insurance payout against the medical bills associated with the given medical record from the insurance company.
\item getdoctor: Takes the address of a doctor and returns all the necessary details associated with that address.
\end{itemize}
\subsubsection{Doctor Functionalities}
\begin{lstlisting}[language=Solidity]
function getrecord(address _p, uint _rid) public checkde(msg.sender) view returns(string memory _rec ){
    return(patient_list[_p].recordlist[_rid].pres);
}
function getpatient(address _p) public checkde(msg.sender) view returns(uint256 _age, string memory _gender){
    return(patient_list[_p].age,
    patient_list[_p].gender);
}
function getrecordnum (address _p) public checkde(msg.sender) view returns(uint256 _rn){
    return patient_list[_p].recn;
}
function addpres( address _p, string memory _pres) public checkde(msg.sender) recordexist(_p,msg.sender){
    patient_list[_p].recordlist[patient_list[_p].recn].pres=_pres;
}
\end{lstlisting}
\begin{itemize}
\item getrecord: Takes the patient's address and record number and returns the respective medical record of that patient.
\item getpatient: Takes the patient's address and returns the information regarding the patient essential for medical diagnosis.
\item getrecordnum: Takes the patient's address and returns the total number of medical records associated with the patient.
\item addpres: Takes the patient's address and sets the prescription in the current medical form associated with the patient with the help of IPFS hash value.

\end{itemize}
\subsubsection{Hospital Functionalities}
\begin{lstlisting}[language=Solidity]
function addrecord(address _p,address _d,uint256 _adm, uint256 _dis ) public hospitalexists(msg.sender){
    patient_list[_p].recn++;
    patient_list[_p].recordlist[patient_list[_p].recn].hospital=msg.sender;
    patient_list[_p].recordlist[patient_list[_p].recn].doctor=_d;
    patient_list[_p].recordlist[patient_list[_p].recn].admissionDate=_adm;
    patient_list[_p].recordlist[patient_list[_p].recn].dischargeDate=_dis;

}
\end{lstlisting}
\begin{itemize}
\item addrecord: Takes the patient's address and all other relevant information and creates a new medical record associated with that patient using these pieces of information.
\end{itemize}
\subsubsection{Insurance Company Functionalities}
\begin{lstlisting}[language=Solidity]
function addcustomer(address _c) public icompexists(msg.sender){
    insurancec_list[msg.sender].iscustomer[_c]=true;
    isinsured[_c]=true;
} 
function removecustomer(address _c) public icustomer(msg.sender, _c){
    insurancec_list[msg.sender].iscustomer[_c]=false;
    isinsured[_c]=false;
} 
function getrecordi(address _c) public payrequire (msg.sender,_c) view returns(string memory _pres, string memory _bill) {
    
    return(
           patient_list[_c].recordlist[ilist[_c][msg.sender]].pres,
           patient_list[_c].recordlist[ilist[_c][msg.sender]].bill 
    );
}
function inspayment(address _c) public payrequire (msg.sender,_c) payable{
      (bool status,)=_c.call{value:msg.value}("");
      require(status,"Transaction Unsucessful");
      insurancec_list[msg.sender].flagraised[_c]=false;
}
\end{lstlisting}
\begin{itemize}
\item addcustomer: Takes the customer’s address and sets the respective customer’s status as insured and also appends it in the list of its customer.
\item removecustomer: Takes the customer’s address and reset the customer’s insurance status with the associated company.
\item getrecordi: Takes the customer’s address and returns the medical prescription and bills associated with the medical record against which payout is requested by the customer.
\item inspayment: Takes the customer’s address and disperses the requested payout by the customer while resetting the request status.
\end{itemize}
\subsubsection{Pharmacy Functionalities}
\begin{lstlisting}[language=Solidity]
function getrecordp(address _c) public isall(_c,msg.sender) view returns(string memory _pres) {
    return(
        patient_list[_c].recordlist[permlist[_c][msg.sender].recid].pres
    );
}
function setbill(address _c, string memory _bill) public isall(_c,msg.sender){
    patient_list[_c].recordlist[permlist[_c][msg.sender].recid].bill=_bill;
    permlist[_c][msg.sender].all=false;
}
\end{lstlisting}
\begin{itemize}
\item getrecordp: Takes the customer’s address and returns the specific prescription associated with the customer's medical record, which is allowed to be accessed by the pharmacy.
\item setbill: Takes the customer’s address and sets the medical bill in the respective medical record of the customer as a form of the hash value.   
\end{itemize}
\subsubsection{Modifiers}
Modifiers is a code used for automatically checking a condition, prior to executing a function, thus it changes the behavior of the function in a declarative way.
\begin{lstlisting}[language=Solidity]
modifier checkp(address _p){
    require(!ispatient[_p],"Already Registered");
    _;
}
modifier checkpe(address _p){
   require(ispatient[_p],"Not Registered");
    _; 
}
modifier recexist(address _p, uint256 n)
{
    require(n<=patient_list[msg.sender].recn,"Not Valid");
    _;
}
modifier allowins(address _i)
{
    require(isinsured[_i],"Don't Have Insurance");
    _;
}
modifier checkd(address _d){
    require(!isdoctor[_d],"Already Registered");
    _;
}
modifier checkde(address _d){
    require(isdoctor[_d],"Not Registered");
    _;
}
modifier hospitalreg(address _h){
    require(!reghospital[_h],"Already Registered");
    _;
}
modifier hospitalexists(address _h){
    require(reghospital[_h],"Not Registered");
    _;
}
modifier recordexist(address _p, address _d){
    require(patient_list[_p].recordlist[patient_list[_p].recn].doctor==_d,"Record Don't Exist");
    _;
}
modifier icompreg(address _i){
    require(!reginsurance[_i],"Already Registered");
    _;
}
modifier icompexists(address _i){
    require(reginsurance[_i],"Not Registered");
    _;
}
modifier icustomer(address _i, address _c){
    require(insurancec_list[_i].iscustomer[_c]=true,"Not a customer");
    _;
}
modifier payrequire(address _i, address _c){
    require(insurancec_list[_i].flagraised[_c], "Request Not Raised");
    _;
}
modifier phcompreg(address _i){
    require(!regpharmacy[_i],"Already Registered");
    _;
}
modifier phcompexists(address _i){
    require(regpharmacy[_i],"Not Registered");
    _;
}
modifier isall(address _c, address _ph)
{
    require(permlist[_c][_ph].all,"Not Allowed");
    _;
}
\end{lstlisting}
\begin{itemize}
\item checkp modifier: This modifier is used to restrict function in such a manner that if a patient is not registered in the network, then the function is accessible. 
\item checkpe modifier: This modifier is used to restrict function in such a manner that if a patient is registered in the network, then the function is accessible. 
\item recexist modifier: This modifier is used to restrict function in such a manner that a patient or doctor can access the information which the function returns if and only if the given record number exists and is valid.
\item checkd modifier: This modifier is used to restrict function in such a manner that if a doctor is not registered in the network, then the function is accessible.
\item checkde modifier: This modifier is used to restrict function in such a manner that if a doctor is registered in the network, then the function is accessible.
\item hospitalreg modifier: This modifier is used to restrict function in such a manner that if a hospital is not registered in the network, then the function is accessible.
\item hospitalexists modifier: This modifier is used to restrict function in such a manner that if a hospital is registered in the network, then the function is accessible.
\item recordexist modifier: This modifier is used to restrict function in such a manner that a doctor can access the information and operation which the function returns and perform respectively if and only if the same doctor is associated with the given patient record number.
\item icompreg modifier: This modifier is used to restrict function in such a manner that if an insurance company is not registered in the network, then the function is accessible.
\item icompexists modifier: This modifier is used to restrict function in such a manner that if an insurance company is registered in the network, then the function is accessible.
\item icustomer modifier: This modifier is used to restrict function in such a manner that if the given patient is a customer of the given insurance company, then the function is accessible.
\item payrequire modifier: This modifier is used to restrict function in such a manner that if the given patient has raised an insurance claim from the  insurance company, then the function is accessible.
\item phcompreg modifier: This modifier is used to restrict function in such a manner that if a pharmacy is not registered in the network, then the function is accessible. 
\item phcompexists modifier: This modifier is used to restrict function in such a manner that if a pharmacy is registered in the network, then the function is accessible.
\item isall modifier: This modifier is used to restrict function in such a manner that if the given pharmacy is allowed to access the medical prescription of the given patient, then the function is accessible.
\end{itemize}
\subsection{Security Analysis}
In this subsection, we will analyze the security of the proposed blockchain-based healthcare system.
\begin{enumerate}
  \item \textbf{Privacy and Confidentiality: }Medical records of an individual are very sensitive information and are prone to risks and vulnerabilities like unauthorized access, sharing, processing, and disclosure. In the proposed model, we have defined a set of rules and regulations using modifiers to ensure that only authorized stakeholders can see patient data and medical information. Besides this, the information sharing among the stakeholders is based on the principle of "Need To Know."
  \item \textbf{Sybil Attack: }Sybil Attacks \cite{21} are a type of attack where an attacker subverts the network by creating a large number of pseudonymous nodes in the network. As in our proposed solution, only authorized or permissioned nodes can create new nodes in the network; thus, no individual has the right to create new nodes. 
  \item \textbf{DDoS: }Distributed denial-of-service is a malicious attack that disrupts a network's regular traffic or availability by overwhelming target infrastructure with a flood of traffic. As the proposed system is decentralized, the attacker must perform DDoS to every single node in the network, which is not feasible, and a byzantine fault-tolerant system helps in locating the failed or malicious nodes in the network. 
\end{enumerate}
The security and performance of the system can also be improved with the help of an ideal consensus algorithm based on the proposed solution's application. For instance, the use of consortium-based blockchain and PoA (Proof of Authority) as consensus algorithm is ideal for the proposed solution.

\section{Conclusion}
The idea of adapting blockchain technology and applying decentralization principles in the healthcare sector for effective data management and handling complex medical services is a compelling one, as healthcare is an important domain for our society. In this paper, we have defined the problematic healthcare processes and demonstrated the use of blockchain-based smart contracts for the medical records handling, effective sharing, accessibility, interoperability, and auditability. Based on the blockchain technology, this system allows patients to share their medical records safely with other stakeholders-all while keeping complete control over their medical data. The use of decentralized architecture ensures the security, privacy, availability, and access control of the medical records. As highlighted in the paper, the focal objectives of the proposed solution are reducing the costs of healthcare services, increasing resource utilization, and thus creating an efficient patient-centric ecosystem. In the current healthcare system, blockchain can help in many ways; reduce administrative and cognitive burdens, reduce enforcement costs and remove intermediaries. Finally, we can conclude that the proposed blockchain-based solution is scalable, iterative, secure, and accessible and will solve many of the current healthcare system’s issues.
As a next step, we intend to expand the current solution by integrating it with other healthcare-related supply chains and services. Also, we intend to expand the data sharing by introducing an incentivization system that will incentivize the patient to support the system for medical researchers. Additionally, we also intend to evaluate our solution on various blockchain networks and consensus algorithms.

\bibliographystyle{unsrt}  
\bibliography{references}  

\end{document}